\documentclass[12pt]{elsarticle}
\biboptions{sort&compress}
\usepackage[T1]{fontenc}
\usepackage[utf8]{inputenc}
\usepackage{microtype}
\usepackage{adjustbox}
\usepackage{amsmath,amssymb,graphicx,booktabs,siunitx,url,subcaption,tabularx}
\usepackage[table]{xcolor}
\usepackage{placeins}
\usepackage{float}
\usepackage{tikz}
\usetikzlibrary{arrows.meta,positioning,fit,shapes.geometric,calc}
\usepackage[colorlinks=true,linkcolor=blue,citecolor=blue,urlcolor=blue,filecolor=blue]{hyperref}
\usepackage{doi}
\makeatletter
\def\ps@pprintTitle{%
  \let\@oddhead\@empty
  \let\@evenhead\@empty
  \let\@oddfoot\@empty
  \let\@evenfoot\@oddfoot}
\makeatother
\graphicspath{{figures/}}
\newcommand{\fslip}{f_{\mathrm{slip}}}
\newcommand{\Leq}{L_{\mathrm{c}}}
\newcommand{\Sh}{\mathit{Sh}}
\newcommand{\gapcell}[2]{\parbox[t]{#1}{\raggedright #2\par}}

\begin{document}
\begin{frontmatter}
\title{\texorpdfstring{\vspace*{-20mm}}{}Design and control of segmented monolith contactors for atmospheric water capture: a gas-side transport strategy tested with CFD}
\author[ntnu]{Patrick Lima\corref{cor}}
\author[ntnu]{Svend Markus Schultz Eeg}
\author[porto]{Margarida S.C.A. Brito}
\author[porto]{Alexandre Ferreira}
\author[ntnu]{Idelfonso Nogueira}
\cortext[cor]{Corresponding author. Email: patrick.lima@ntnu.no}
\address[ntnu]{Norwegian University of Science and Technology, Trondheim, Norway}
\address[porto]{LSRE-LCM, ALiCE, Faculty of Engineering, University of Porto, Rua Dr. Roberto Frias, 4200-465 Porto, Portugal}
\begin{abstract}
\footnotesize
Atmospheric moisture is a candidate local water source, but contactors must remove vapour from dilute, humidity-varying air at low pressure loss, and designing them with 3D CFD is costly. This study turns a few CFD runs per geometry into a workflow for the design, operation and control of segmented monolith contactors, worked out at the gas-side limit of a wall perfectly permeable to water vapour, which bounds any real sorbent. A capture law with an effective Sherwood number and a length correction, fitted to the CFD runs of each geometry, chooses geometry and length; two runs per geometry already calibrate it to a median error of 2.8 mm in length. Seven designs predicted for 99.9\% capture and simulated with new CFD at lengths fixed in advance captured 99.896--99.902\%. In a laboratory-scale module the fixed losses of ports and plenums make length, not pressure, the deciding variable: the nondominated designs span a factor of two to three in length for at most 27\% more pressure, and offset channels give the shortest unit, 73 mm against 142 mm for the lowest-pressure design. The captured rate, not the efficiency, follows humidity and is the variable to regulate, with the airflow as handle. On a planar CFD section of a near-optimal geometry, a PI controller tuned from one step test holds the rate, and humidity feedforward lowers the integrated error by a further 70\textendash{}72\%.
\end{abstract}
\end{frontmatter}
\clearpage

\section{Introduction}
Reliable access to drinking water remains uneven. In 2024, approximately
2.1 billion people lacked safely managed drinking-water services; coverage
was 60\% in rural areas and 83\% in urban areas \cite{WHO_UNICEF2025JMP}.
These disparities motivate research on complementary local water sources
for communities where conventional supply is constrained. Atmospheric moisture is a possible local source of feed water for such systems. Sorbent-based
atmospheric water harvesting can capture vapour even at low relative
humidity, as shown by a device operated in an arid climate
\cite{Kim2018}. Reviews stress that deployment depends on turning sorbent performance into working devices and processes \cite{Guo2023,Bai2024,Borne2025}.

Turning atmospheric moisture into a dependable water supply is a coupled materials, device and process problem. A sorbent must take up water at the relevant humidity and adsorb and desorb it quickly, and regeneration and condensation need effective heat and mass transfer \cite{Bai2024}. At device scale, large volumes of dilute air of varying humidity must reach the sorbent without excessive airflow resistance or equipment size \cite{Borne2025}. Structured monolith and fibre contactors are designed for this, but bring geometric, kinetic and thermal trade-offs \cite{DeWitt2018,Rezaei2010}, and uneven flow among channels changes their apparent transfer coefficients \cite{Sharma2020}. Transient three-dimensional models of adsorbent beds have coupled internal and external heat and mass transfer \cite{Wang2022Honeycomb}. At system scale, weather affects capture and regeneration, so scheduling and operation that follow the climate matter beyond the nominal design \cite{Kim2024Climate,Almassad2022}, as they do for solid-sorbent direct air capture \cite{Wiegner2022}.

Table~\ref{tab:intro-gap} groups this literature into four levels and the question that each leaves open. This work addresses the step between them: a design suggested by a reduced model must be checked against its capture target before it is adopted, and the operating target must follow the vapour available at the current humidity.

\begin{table}[!ht]
\centering\footnotesize
\caption{Positioning of the present study relative to the literature reviewed here.}
\label{tab:intro-gap}
\setlength{\tabcolsep}{3pt}
\begin{adjustbox}{max width=\textwidth}\begin{tabular}{@{}llll@{}}
\toprule
\gapcell{.17\textwidth}{Literature level} & \gapcell{.29\textwidth}{Representative focus} & \gapcell{.25\textwidth}{Open question} & \gapcell{.25\textwidth}{Role in this study} \\
\midrule
\gapcell{.17\textwidth}{Materials and devices} & \gapcell{.29\textwidth}{Sorbent uptake, cycling and harvesting devices \cite{Kim2018,Bai2024}} & \gapcell{.25\textwidth}{From material potential to air-side contactor design} & \gapcell{.25\textwidth}{Fresh-surface transport limit, separate from cyclic yield}\\[0.6mm]
\gapcell{.17\textwidth}{Structured contactors} & \gapcell{.29\textwidth}{Monolith and fibre architecture, coupled transport \cite{DeWitt2018,Wang2022Honeycomb}} & \gapcell{.25\textwidth}{Geometry screening under capture and pressure constraints} & \gapcell{.25\textwidth}{Segmented 3D geometries, capture law, new CFD checks}\\[0.6mm]
\gapcell{.17\textwidth}{Climate-aware operation} & \gapcell{.29\textwidth}{Kinetics, scheduling, adaptive cycle timing \cite{Kim2024Climate,Almassad2022}} & \gapcell{.25\textwidth}{Choice of controlled and manipulated variables} & \gapcell{.25\textwidth}{Captured mass rate as controlled variable, airflow as manipulated variable}\\[0.6mm]
\gapcell{.17\textwidth}{CFD-based control} & \gapcell{.29\textwidth}{Virtual-plant feedback structures \cite{Desta2004,Crose2019,Lima2025Control}} & \gapcell{.25\textwidth}{Controllability at the design stage} & \gapcell{.25\textwidth}{Controllability check on a planar CFD plant} \\
\bottomrule
\end{tabular}\end{adjustbox}
\end{table}

Computational fluid dynamics (CFD) resolves the flow and vapour fields that set these trade-offs and can test a design before it is built. Compartmental, reduced-order and surrogate models derived from CFD have supported process decisions \cite{Yang2019,Lang2011,Bezzo2003,Bhosekar2018}, and CFD has supported geometry and operating-condition optimisation \cite{Lima2026Fuzzy,Lima2026Rineng,Lima2026CACEBO}. It has also served as the virtual plant for controller development, from ventilation and thin-film deposition to fluidised beds and core--annular flow \cite{Desta2004,Crose2019,LiShen2022,Lima2025Control,Lima2025ComparativeControl}; in adsorption, model-based feedforward and feedback have been combined for pressure-swing tracking \cite{Bitzer2005}. For a water-capture contactor, CFD can link local vapour transport to capture, pressure loss and an operating variable.

Equipment design decisions, such as the geometry and length of a contactor, are seldom tested for controllability, the ability of the physical unit to hold its target with bounded inputs while disturbances act \cite{SkogestadPostlethwaite2005}. Process systems engineering has nevertheless long argued for designing process and control together \cite{Sakizlis2004}. For a contactor designed with CFD, such a test needs transient simulations of the whole device for identification and for repeated closed-loop runs, which are costly in 3D. Here the controllability check is therefore made on a planar CFD plant, a two-dimensional longitudinal section of the $5\times6$--O geometry, one of the near-optimal designs of the screening.

\begin{figure}[H]
\centering\includegraphics[width=1\linewidth]{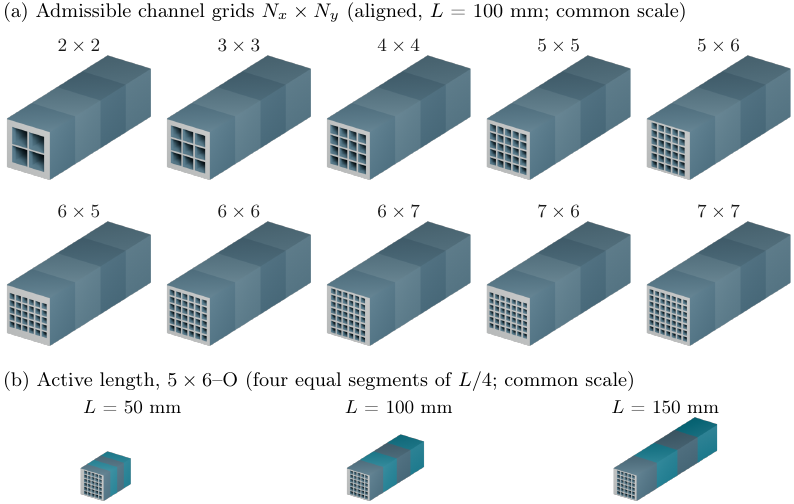}
\caption{The design set at a common scale. (a) Aligned square-channel grids in the
same transverse envelope, from few large channels to many small ones. (b)
Active-length variants of the $5\times6$ offset architecture (50, 100 and 150 mm).
Colours in (b) identify the
lateral positions A and B. Ports and plenums are omitted.}
\label{fig:catalogue}
\end{figure}

The contactors studied form one parametric family of segmented monoliths, the design set (Fig.~\ref{fig:catalogue}). All members share a $30\times30$ mm transverse envelope, the size of a laboratory module, and 1 mm internal walls, so more channels means smaller channels; the active length also varies. A member is named by its channel grid $N_x\times N_y$ and its architecture, aligned (A) or offset (O), for example $5\times6$--O.

\begin{figure}[!ht]
\centering\includegraphics[width=\linewidth]{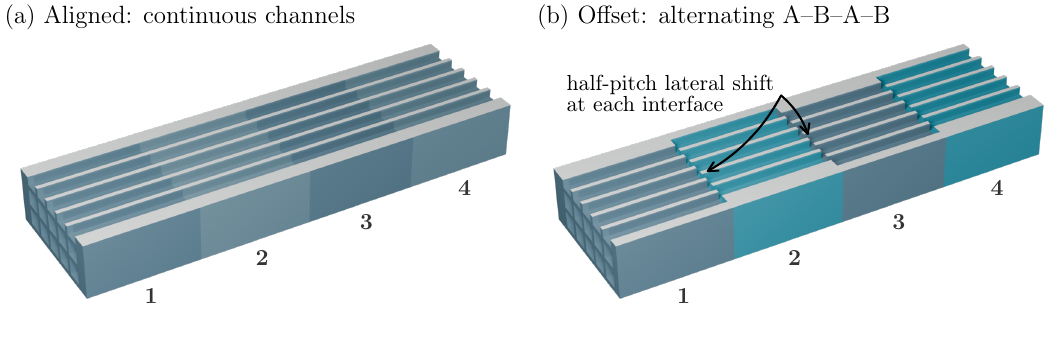}
\caption{Concept of the two architectures, shown for the $5\times6$ grid. (a) Aligned
(A): continuous channel positions. (b) Offset (O): alternate segments displaced by half a
channel pitch, with lateral positions A and B in colour. Cutaway renders of the CFD
geometry; numbers mark the axial segments.}
\label{fig:concept}
\end{figure}

In the aligned architecture (A) the channel positions are continuous across the four axial segments. In the offset architecture (O) alternate segments are displaced by half a pitch, so the flow enters new passages at each segment interface, where the concentration layer can start again. The price is additional pressure drop (Fig.~\ref{fig:concept}). Offset honeycomb sections have been described before in fluid purification \cite{Obee2011}.

The objective of this work is a modelling strategy for the design, operation and control of such contactors, built from established models rather than new algorithms. To test it, the sorbent is represented by a limiting case: a wall perfectly permeable to water vapour, an infinite sink with zero vapour mass fraction at the wall. The CFD results then give the gas-side limit of capture, an upper bound for any real sorbent. A finite-capacity sorbent replaces this wall condition and makes capture change with time, and the same steps can then be calibrated on its simulations or measurements. The study delivers the gas-side information a designer needs first: the length and pressure a capture target requires at best, the geometry, and the variable to regulate.

The strategy has three steps: calibrate a capture law, with an effective Sherwood number and a length correction, on a few CFD runs per geometry (or on measurements); use the law to choose geometry and length under capture and pressure constraints; and test the choice with new CFD. The design set is the worked example of these steps. Two further parts use the same CFD model. The operation study determines how airflow and humidity affect capture efficiency and captured mass rate, to identify the controlled variable. The control study tunes a PI controller of the captured rate from one step test on the planar CFD plant, checks it on a set-point change, and tests the rejection of humidity steps with and without humidity feedforward. Two engineering results stand out: in a laboratory-scale module the fixed losses of ports and plenums make length, not pressure, the deciding design variable, and the offset architecture gives the shortest designs.

\FloatBarrier
\section{Methodology}
\label{sec:device}
\begin{figure}[H]
\centering
\includegraphics[width=\linewidth]{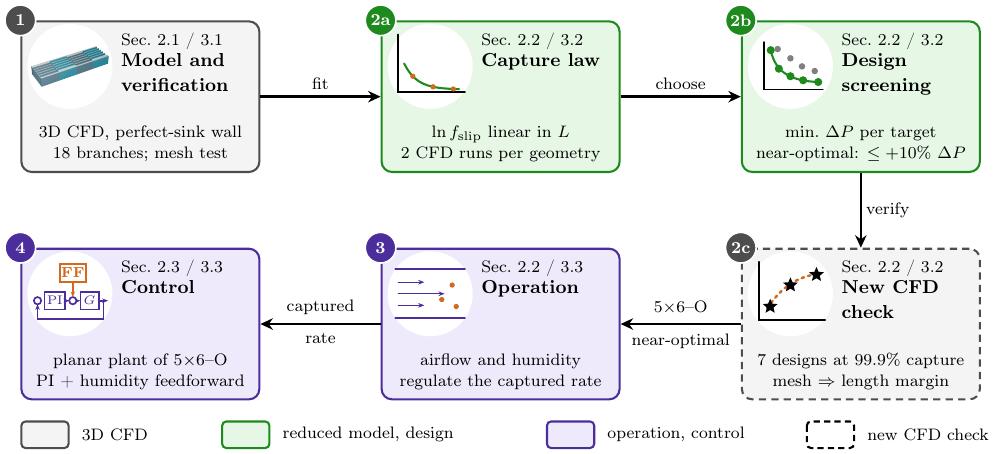}
\caption{Workflow of the study. The tag on each box gives the sections of its methods and results.}
\label{fig:workflow}
\end{figure}

All three decisions use the same wall, perfectly permeable to water vapour (an infinite sink, $Y_{\mathrm{H_2O},w}=0$), so the CFD simulations in this work consider a gas-side limit; a finite-capacity sorbent would replace this boundary condition and make capture time-dependent. Figure~\ref{fig:workflow} organises the work as four stages,
each with its methods and its results. The CFD model and its numerical verification
(Sections~\ref{sec:geometry-mesh} and~\ref{sec:verification}) give the evidence base.
Design uses converged steady 3D simulations, reduced responses and prospective checks to
screen lengths (Sections~\ref{sec:design-method} and~\ref{sec:design-results}). Operation
uses the velocity and humidity dependence of capture efficiency and captured mass rate
to select the controlled variable (Sections~\ref{sec:design-method} and~\ref{sec:velocity}).

Control uses a transient planar CFD plant with PI feedback of
the captured rate $\dot m^{\prime}_c$ against its reference $r$, the inlet velocity $U$
as manipulated variable and a humidity feedforward (Sections~\ref{sec:control-method}
and~\ref{sec:2d-limits}).

\FloatBarrier
\subsection{CFD model and numerical verification}
\label{sec:geometry-mesh}
\label{sec:geometry}
The four-segment aligned and offset architectures are introduced in
Fig.~\ref{fig:concept}. Four equal axial segments follow either continuous
channel positions or an A--B--A--B lateral offset, displaced by half a pitch
without intermediate chambers. Table~\ref{tab:model-at-glance} defines the geometric inputs and fixed
dimensions; Fig.~\ref{fig:catalogue} shows the resulting channel grids and
active lengths at a common scale.

\begin{table}[!ht]
\centering\small
\caption{Geometric inputs and fixed dimensions of the design set. Here $a$ is the internal side length of a square channel.}
\label{tab:model-at-glance}
\begin{tabular}{@{}ll@{}}
\toprule
\gapcell{.22\textwidth}{Item} & \gapcell{.66\textwidth}{Specification} \\
\midrule
\gapcell{.22\textwidth}{Active geometry} & \gapcell{.66\textwidth}{$30\times30$ mm envelope; 1 mm internal walls and outer margin; $a$ is the largest square side that fits the envelope with room for the half-pitch offset, rounded down to 0.05 mm, so $a$ falls from 5.4 mm ($4\times4$) to 2.85 mm ($7\times7$); four equal axial segments} \\[1.2mm]
\gapcell{.22\textwidth}{Design variables} & \gapcell{.66\textwidth}{$[N_x,N_y,\omega,L]$: channel counts, aligned/offset architecture $\omega$ (A or O) and active length; $12a\leq L\leq150$ mm} \\[1.2mm]
\gapcell{.22\textwidth}{Design set} & \gapcell{.66\textwidth}{10 admissible grids; 18 branches (10 offset and 8 aligned; the aligned set retains one representative per unordered pair of channel counts); 14 retained in the nominal fits} \\[1.2mm]
\gapcell{.22\textwidth}{Ports and plenums} & \gapcell{.66\textwidth}{$10\times10$ mm port; 8 mm inlet port and plenum; 8 mm outlet plenum; 30 mm outlet extension} \\[1.2mm]
\bottomrule
\end{tabular}
\end{table}

The internal side of each square channel is $a$, which is also its hydraulic diameter, $D_h=a$. Each of the four segments is at least three hydraulic diameters long, $L\geq12a$. Short segments keep the flow in the concentration entrance region, where the local Sherwood number exceeds its fully developed value \cite{Gundlapally2011}; this is the effect that segmentation and offset are meant to exploit. A branch is a fixed combination $(N_x,N_y,\omega)$ of the channel counts in the two transverse directions and the architecture $\omega$ (A or O), with length free to vary, so the design vector is $\mathbf{x}=[N_x,N_y,\omega,L]$: three discrete choices and one continuous length. The design responses are the capture efficiency $\eta_c$ and the pressure drop $\Delta P$ (Section~\ref{sec:design-method}). Inlet velocity and relative humidity are operating variables, not design variables. Velocities are port velocities unless stated: the nominal 2 m s$^{-1}$ through the $10\times10$ mm port gives channel velocities of about 0.4--0.6 m s$^{-1}$.

\label{sec:physical-model}
The flow is the steady, laminar and isothermal flow of an ideal-gas mixture of water vapour, oxygen and nitrogen, with dilute-mixture diffusion of the vapour; Table~\ref{tab:physical-model} lists the settings. Adsorbent walls impose $Y_{\mathrm{H_2O},w}=0$ and inert walls zero vapour flux, so capture is driven by the concentration gradient towards the sink rather than by a prescribed uptake rate. The integrated wall uptake,
\begin{equation}
 \dot m_{v,w} =\int_{A_{\mathrm{ads}}}\mathbf{J}_v\cdot\mathbf{n}\,\mathrm{d}A,
 \label{eq:wall-uptake-cfd}
\end{equation}
with $\mathbf{J}_v=-\rho D\nabla Y_v$ the diffusive vapour flux, $\rho$ the mixture density, $Y_v$ the vapour mass fraction, $\mathbf{n}$ pointing out of the fluid and $A_{\mathrm{ads}}$ the adsorbent area, gives a species-balance check against the inlet--outlet vapour deficit.

\begin{table}[!ht]
\centering\small
\caption{Physical model, boundary conditions and solver configuration for the steady three-dimensional study.}
\label{tab:physical-model}
\begin{tabular}{@{}ll@{}}
\toprule
\gapcell{.22\textwidth}{Item} & \gapcell{.66\textwidth}{Specification} \\
\midrule
\gapcell{.22\textwidth}{Nominal inlet} & \gapcell{.66\textwidth}{Velocity inlet, 2 m s$^{-1}$, uniform and normal to the boundary; 300 K, 101325 Pa, 50\% relative humidity ($Y_{\mathrm{H_2O}}=0.0109$); pressure outlet at zero gauge pressure} \\[1.2mm]
\gapcell{.22\textwidth}{Air properties} & \gapcell{.66\textwidth}{Ideal-gas density; $\mu=1.849\times10^{-5}$ Pa s; $D_{\mathrm{H_2O-air}}=2.50\times10^{-5}$ m$^2$ s$^{-1}$, written $D$ in the text} \\[1.2mm]
\gapcell{.22\textwidth}{Walls} & \gapcell{.66\textwidth}{No-slip, 300 K; adsorbent $Y_{\mathrm{H_2O},w}=0$, inert walls with zero water flux; the energy equation is kept as an isothermal check (maximum deviation 0.01 K)} \\[1.2mm]
\gapcell{.22\textwidth}{Solver} & \gapcell{.66\textwidth}{Fluent 2024 R2, double precision; SIMPLE pressure--velocity coupling; second-order pressure interpolation; second-order upwind for the transport equations} \\[1.2mm]
\gapcell{.22\textwidth}{Mesh} & \gapcell{.66\textwidth}{Unstructured tetrahedra from one automated recipe for all branches and lengths, set by the maximum element size $h_{\max}$; design mesh $h_{\max}=0.528$ mm} \\
\bottomrule
\end{tabular}
\end{table}

A run counts as converged when its monitors, residuals and balances meet the limits of Table~\ref{tab:acceptance}.

\begin{table}[!ht]
\centering\small
\caption{Convergence criteria of a CFD run. Residuals are maxima over the last 25 iterations; balances are relative to the inlet flow.}
\label{tab:acceptance}
\begin{adjustbox}{max width=\textwidth}\begin{tabular}{@{}ll@{}}\toprule
Check & Limit\\\midrule
Pressure drop, slip, inlet mass flow & stable over two consecutive five-sample windows\\
Residuals: continuity; velocity; energy & $5\times10^{-3}$; $10^{-4}$; $10^{-6}$\\
Residuals: H$_2$O; O$_2$ & $3\times10^{-4}$; $2\times10^{-4}$\\
Total mass balance & $\leq10^{-4}$\\
H$_2$O closure, wall included & $\leq2\times10^{-4}$\\
Outlet & no reversed flow\\
\bottomrule\end{tabular}\end{adjustbox}\end{table}

\begin{figure}[!ht]\centering
\includegraphics[width=.85\linewidth]{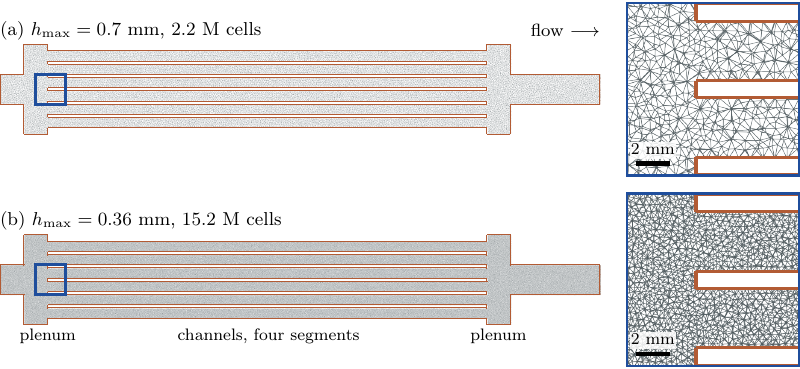}
\caption{Meshes of the adjusted $6\times6$--A design (147 mm) in a longitudinal cut through one channel row: (a) $h_{\max}=0.7$ mm and (b) the finest level, 0.36 mm. Red lines are the channel walls; the blue box marks the zoom on the right. Cell counts refer to the whole domain.}
\label{fig:mesh-visual}\end{figure}

\label{sec:mesh}
Because one recipe meshes every geometry variant, differences between designs do not come from the meshing. Mesh verification uses four representative geometries, the $5\times6$ and $7\times7$ grids in both architectures (57 and 42.75 mm), at five maximum element sizes from 1.056 to 0.4526 mm, and is judged on the quantities the decisions use, capture efficiency and pressure drop, with limits of 0.5 percentage points and 0.5\% between the design mesh and the next finer level. These limits are adequate for screening at 90--99\% capture. At 99.9\% the slip is only 0.1\%, smaller than the 0.5-point limit, so that target gets its own refinement, down to 0.36 mm, on the $6\times6$--A design recommended for it, whose channels are 147 mm long (the adjusted design, Section~\ref{sec:design-results}), whose meshes are shown in Fig.~\ref{fig:mesh-visual}.

\FloatBarrier
\subsection{Design and operation studies}
\label{sec:design-method}
\label{sec:optimisation}
\label{sec:design-data}
Table~\ref{tab:groups} lists the simulation groups. Only converged nominal observations enter the fits; the $2\times2$ and $3\times3$ branches did not converge at their minimum lengths, which leaves 14 fitted branches with channel sides from 5.4 to 2.85 mm.

\begin{table}[!ht]
\centering\footnotesize
\caption{Simulation groups and their roles. The four mesh-study geometries also belong to the nominal group.}
\label{tab:groups}
\begin{adjustbox}{max width=\textwidth}\begin{tabular}{lrl}\toprule
Group & Observations & Purpose \\\midrule
Nominal & 48 & Branch-specific response fits \\
Operating velocity & 18 & Throughput--capture comparison \\
Prospective and resolution checks & 17 & 7 target, 5 finer-mesh, 5 adjusted-design runs \\
Mesh refinement & 20 & Four geometries at five resolutions \\
Humidity, 3D & 2 & Adjusted design at RH 20 and 80\% \\
\bottomrule\end{tabular}\end{adjustbox}
\end{table}

The vapour slip $\fslip=|\dot m_{v,\mathrm{out}}|/|\dot m_{v,\mathrm{in}}|$ is the fraction of inlet vapour that leaves uncaptured, the capture efficiency is $\eta_c=1-\fslip$, the captured mass rate $\dot m_{v,\mathrm{capt}}$ is the difference of the inlet and outlet vapour flows, and the pressure drop is the difference of the area-averaged static pressures at the two ports. Results are reported as capture efficiency, but the response model uses the slip, whose logarithm is nearly linear in length. The hydraulic power is $P_h=Q\Delta P$, with $Q$ the inlet volumetric flow; it excludes fan efficiency.

\label{sec:response-method}
The Sherwood number, $\Sh=k_mD_h/D$, compares convective mass transfer, with gas-side coefficient $k_m$, with molecular diffusion. A duct mass balance with a perfect-sink wall, the mass-transfer analogue of a duct at constant wall temperature \cite{Bergman2011}, gives the capture law
\begin{equation}
 \fslip=\exp[-\Sh(4D/Q)N(L+\Leq)],\qquad N=N_xN_y,
 \label{eq:slip}
\end{equation}
where $N$ is the number of channels, $\Sh$ is an effective coefficient of the branch, inferred from integral CFD responses, and the length correction $\Leq$ lumps the enhanced transfer of the entrance region and other end effects into an extra effective length, as is common in monolith channel models \cite{Tronconi1992,Gundlapally2011}. Since $\ln\fslip$ is linear in $L$, its slope gives $\Sh$ and its intercept $\Leq$. Pressure is represented by
\begin{equation}
 \Delta P=\Delta P_0+kL, \label{eq:dp}
\end{equation}
with $\Delta P_0$ a fitted intercept and $k$ the pressure gradient along the active length. Ordinary least squares fits the four constants of each branch, two per response; two lengths determine them, and further lengths test them.

The law is tested by calibration: it is fitted on the shortest one or two lengths of each branch with at least three lengths and predicts the longer ones, which measures how many CFD runs per geometry the law needs.

\label{sec:selection-method}
The design problem has two objectives, the pressure drop and the slip (equivalently, the capture efficiency). We treat it in $\varepsilon$-constraint form: for a given capture target, the pressure drop is minimised subject to a limit on the slip,
\begin{equation}
\begin{aligned}
 \min_{b\in\mathcal B_{\rm fit},\,L}\quad &\widehat{\Delta P}_b(L)\\
 \text{subject to}\quad &\widehat{\fslip}_b(L)\leq f_{\rm target},\\
 &12a_b\leq L\leq150\ \mathrm{mm},
\end{aligned}
\label{eq:design-screen}
\end{equation}
where $b=(N_x,N_y,\omega)$ is one of the 14 fitted branches $\mathcal B_{\rm fit}$, $a_b$ its channel side, hats denote reduced-model predictions and $f_{\rm target}=1-\eta_{\rm target}$. Each target gives the design of lowest predicted pressure, and repeating the problem over several targets traces the predicted Pareto front of nondominated capture--pressure designs. Designs whose predicted pressure is at most 10\% above the lowest form the near-optimal set; we report its shortest member, to expose the trade between length and pressure. A target of 99.9\% is the demanding case study: its recommended lengths are fixed before new CFD checks on the design and finer meshes, which do not enter any fit.

In practice the strategy reduces to four steps for each new geometry or target:
\begin{enumerate}\setlength{\itemsep}{1pt}\setlength{\topsep}{2pt}\setlength{\parsep}{0pt}
\item run the geometry at two lengths and fit the slope and intercept of $\ln\fslip$ against $L$, which give $\Sh$ and $\Leq$ (Eq.~\eqref{eq:slip});
\item fit the pressure drop against $L$ in the same way (Eq.~\eqref{eq:dp});
\item solve Eq.~\eqref{eq:design-screen} for the capture target, which gives the geometry and length of lowest pressure and the shortest design within 10\% of it;
\item test the choice with new CFD at that length and add a margin for the mesh.
\end{enumerate}

Operation is studied with the same steady 3D model and the same
indicators. The velocity study compares $5\times6$--A and --O at 100 mm and $7\times7$--O at 70 mm
at six inlet velocities from 0.8 to 2.6 m s$^{-1}$ (channel Reynolds number about 36--144; 18 simulations, Table~\ref{tab:groups}); capture
efficiency, captured mass rate, pressure drop and hydraulic power are compared. Humidity is varied at
fixed velocity (RH 20, 50 and 80\%) on the planar plant of Section~\ref{sec:control-method} and, as a
check, on the adjusted 3D design. The aim is to decide which quantity should be regulated as humidity
changes.

\FloatBarrier
\subsection{Control model and closed-loop evaluation}
\label{sec:control-method}
\label{sec:control-plant}
The control study needs one transient trajectory for identification and one for each closed-loop scenario. The planar CFD plant, a two-dimensional longitudinal slice of the 3D geometry, makes these runs affordable: it has 34,508 cells against 2.8--5.0 million for the 3D meshes of the designs at 99.9\% capture, and each closed-loop run took 18--53 min on two solver processes, whereas a transient 3D trajectory would need roughly 80--145 times more processor time, since the cost per iteration grows with the number of cells. The check is therefore limited to the planar plant
(Section~\ref{sec:2d-limits}).
The plant is taken from the near-optimal set of the screening rather than from its single lowest-pressure design: at 99.9\% capture the $5\times6$--O design is predicted within 1.3\% of the lowest pressure, and new CFD confirmed its capture (Fig.~\ref{fig:front}c). The plant is a section of this geometry at 100 mm, the length used in the operating study and in the field diagnostics, so its section at $y=12.6$ mm links design, operation and control; it is not the 134 mm unit of the 99.9\% design, but at its operating point it also captures 99.9\%. The section retains the in-plane channels, interfaces, plenums and ports and omits the perpendicular walls and gradients, so it is identified as a plant of its own.

The planar plant solves the transient counterparts of the steady equations of Section~\ref{sec:geometry-mesh}, with the same closures and perfect-sink wall, with second-order implicit time integration and a time step of 0.005 s (15 inner iterations per step in the closed-loop runs).

The controlled output is the integrated adsorbent-wall vapour uptake over the adsorbent-wall boundary $\Gamma_{\mathrm{ads}}$, per unit out-of-plane depth, denoted $\dot m'_c$; it is the planar counterpart of the wall uptake of Eq.~\eqref{eq:wall-uptake-cfd}. With positive flux directed from the fluid into the wall,
\begin{equation}
 \dot m'_c(t)=\int_{\Gamma_{\mathrm{ads}}} j_{v,w}(s,t)\,\mathrm{d}s,
 \qquad [\dot m'_c]=\mathrm{kg\,s^{-1}\,m^{-1}}.
 \label{eq:wall-capture}
\end{equation}
Here $j_{v,w}=-\rho D\,\partial Y_v/\partial n$ is the local normal vapour flux into the wall, and the prime marks a quantity per unit omitted depth. The CFD wall flux is used as an ideal measurement, without sensor noise or delay; a physical unit would estimate the rate from an inlet--outlet vapour balance.

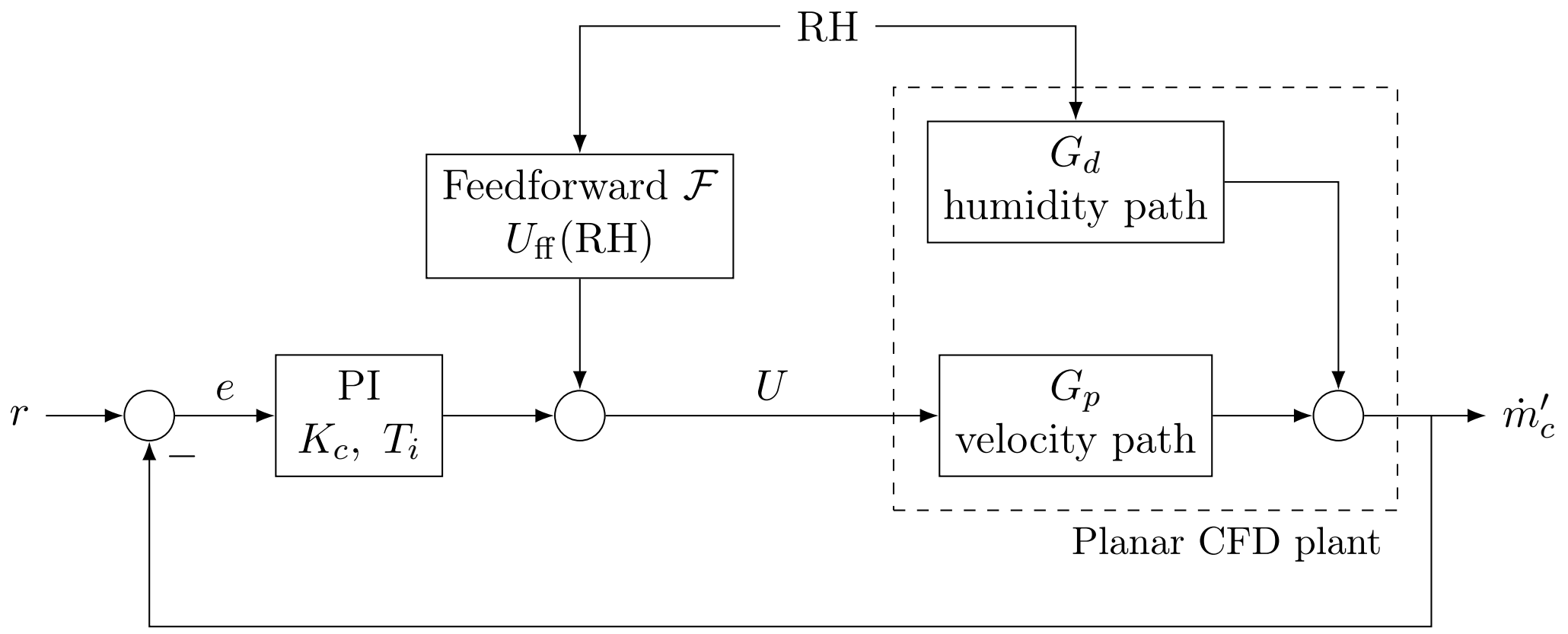
\begin{figure}[!ht]\centering
\begin{adjustbox}{max width=\linewidth}
\begin{tikzpicture}[>=Latex, node distance=6mm and 9mm, every node/.style={font=\small},
  blk/.style={draw, minimum height=9mm, minimum width=15mm, align=center, fill=white},
  sum/.style={draw, circle, inner sep=0pt, minimum size=4.5mm, fill=white}]
\node (r) {$r$};
\node[sum, right=7mm of r] (s1) {};
\node[blk, right=9mm of s1] (pi) {PI\\$K_c,\ T_i$};
\node[sum, right=10mm of pi] (s2) {};
\node[blk, right=30mm of s2] (gp) {$G_p$\\velocity path};
\node[sum, right=9mm of gp] (s3) {};
\node[right=11mm of s3] (y) {$\dot m'_c$};
\node[blk, above=10mm of s2] (ff) {Feedforward $\mathcal F$\\$U_{\mathrm{ff}}(\mathrm{RH})$};
\node[blk, above=10mm of gp] (gd) {$G_d$\\humidity path};
\node (rh) at ($(ff.north)!0.5!(gd.north)+(0,10mm)$) {RH};
\node[draw, dashed, fit=(gd)(gp)(s3), inner sep=3mm, label={[anchor=north east, font=\footnotesize]south east:Planar CFD plant}] (plant) {};
\draw[->] (r) -- (s1);
\draw[->] (s1) -- node[above] {$e$} (pi);
\draw[->] (pi) -- (s2);
\draw[->] (s2) -- node[above] {$U$} (gp);
\draw[->] (gp) -- (s3);
\draw[->] (s3) -- (y);
\draw[->] (ff) -- (s2);
\draw[->] (gd) -| (s3);
\draw[->] (rh) -| (ff);
\draw[->] (rh) -| (gd);
\draw[->] ($(s3.east)!0.55!(y.west)$) -- ++(0,-19mm) -| node[pos=0.96, right] {$-$} (s1.south);
\end{tikzpicture}
\end{adjustbox}
\caption{Control structure: PI feedback of the captured rate with humidity feedforward. Both paths of the plant are computed by the same planar CFD model; $G_p$ and $G_d$ are their identified first-order-plus-delay models.}
\label{fig:control-structure}\end{figure}

The inlet velocity $U$ is the manipulated variable, bounded to $0.2\leq U\leq1.4$ m s$^{-1}$, and the inlet relative humidity (RH) is the disturbance. The control structure is the classical combination of feedback and feedforward (Fig.~\ref{fig:control-structure}). The velocity path $G_p$ and the humidity path $G_d$ of the plant are each described by a first-order-plus-delay model in deviation variables,
\begin{equation}
 G_p(s)=\frac{K_p\,e^{-\theta_p s}}{\tau_p s+1},\qquad G_d(s)=\frac{K_d\,e^{-\theta_d s}}{\tau_d s+1},
 \label{eq:fopdt}
\end{equation}
identified from a 10\% velocity step and from an open-loop humidity step at RH 50\%, both started from the same steady state. The simple internal model control (SIMC) rules \cite{Skogestad2006} give the proportional--integral (PI) controller
\begin{equation}
 K_c=\frac{\tau_p}{K_p(\tau_c+\theta_p)},\qquad
 T_i=\min\{\tau_p,4(\tau_c+\theta_p)\},
 \label{eq:pi-tuning}
\end{equation}
with closed-loop time $\tau_c=\tau_p$. The feedforward is the ideal $-G_d/G_p$ \cite{Seborg2016,Guzman2011}: a delay $\theta_d-\theta_p$ and a lead--lag $(\tau_ps+1)/(\tau_ds+1)$ applied to a steady map $U_{\mathrm{ff}}(\mathrm{RH})$ that gives the velocity holding the rate target at each humidity (Section~\ref{sec:humidity-control}). The command at each controller update $k$, every $T_s=0.010$ s, is
\begin{equation}
 U_k=\mathrm{clip}\!\left(K_c e_k+I_k+\mathcal F\!\left[U_{\mathrm{ff}}(\mathrm{RH})-U_{\mathrm{ff}}(50\%)\right]_k,\,0.2,\,1.4\right),
 \label{eq:pi-ff}
\end{equation}
where $e_k=r_k-\dot m'_{c,k}$, $I_k$ is the integral term with conditional integration against windup \cite{Astrom1989} and $\mathcal F$ is the delay and lead--lag filter. Without feedforward the last term is zero. The identified parameters are given in Section~\ref{sec:2d-limits}.

\label{sec:control-evaluation}
The humidity steps RH 50$\to$30$\to$70$\to$50\% at 0.4, 1.4 and 2.4 s are applied with the rate reference fixed at 0.053 g s$^{-1}$ m$^{-1}$, at fixed velocity, with PI and with PI plus feedforward. Performance is measured by the integrated absolute error, $\mathrm{IAE}=\int|r-\dot m'_c|\,\mathrm{d}t$ from 0.4 s, in mg m$^{-1}$. A second humidity sequence, with all parameters frozen, tests the feedforward away from the data used to identify it.

\FloatBarrier
\section{Results and discussion}
\subsection{Numerical verification}
\label{sec:verification}
Mesh refinement changes the capture by a fraction of a percentage point, which is small in absolute terms but large in slip, and it moves the length needed for 99.9\% capture by a few millimetres.

\begin{figure}[H]\centering
\includegraphics[width=\linewidth]{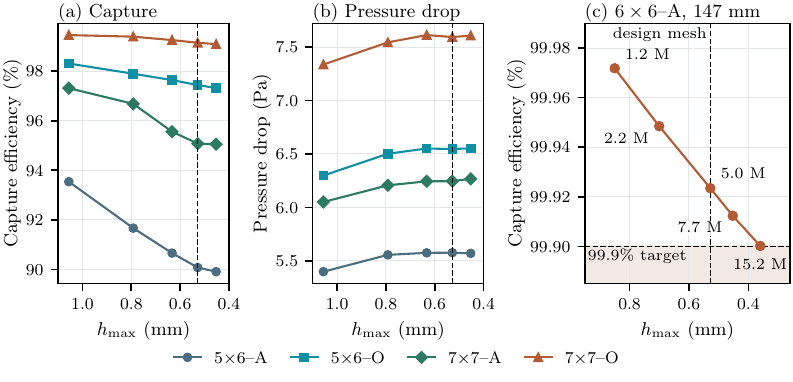}
\caption{Mesh test: (a) capture and (b) pressure drop of the four test geometries; (c) capture of the adjusted $6\times6$--A design, labelled with the cells of the whole domain. Dashed lines mark the design mesh and the 99.9\% target.}
\label{fig:mesh-test}\end{figure}

Between the design mesh, $h_{\max}=0.528$ mm, and the finer 0.4526 mm level, the capture of the four screening geometries changes by $-0.168$ to $-0.022$ percentage points and the pressure drop by at most 0.351\%, within the limits of 0.5 percentage points and 0.5\% (Fig.~\ref{fig:mesh-test}a,b). The design mesh is therefore adequate for screening at 90--99\% capture.

The 99.9\% target asks more of the discretisation, so the adjusted $6\times6$--A design was refined at five element sizes down to 0.36 mm (Fig.~\ref{fig:mesh-test}c). Its capture changes by 0.023 percentage points, from 99.923\% to 99.900\%, and its pressure drop by less than 0.05 Pa. Near 99.9\% even a small change in capture is a large fraction of the slip, so differences are given in percentage points.

In engineering terms the change is a few millimetres of length: converted with the capture law (Eq.~\eqref{eq:slip}), the step from the design mesh to 0.36 mm equals about 5.5 mm of channel, close to the 4.8 mm margin of the adjusted design (Section~\ref{sec:design-results}), while it costs three times the cells. The capture is still falling at 0.36 mm, and extrapolated to finer meshes the full effect is at worst about 1 cm, 7\% of the length. At fixed length $-\ln\fslip$, and with it the effective Sherwood number, changes by about 4\%, against a spread of about 40\% (2.2--3.1) among the branches (Section~\ref{sec:law}), so the mesh does not change which geometry is chosen, only its length.

The capture decreases steadily with refinement (Fig.~\ref{fig:mesh-test}c) because the last tenth of a per cent of the vapour reaches the wall only by diffusion across the channel, to which numerical diffusion adds on a coarser mesh. The constants $\Sh$ and $\Leq$ are therefore effective constants of the mesh on which the design is decided, and the mesh effect is carried as a length margin of a few millimetres, sized here on the 0.4526 mm mesh and checked at 0.36 mm. This keeps every design decision on the inexpensive design mesh.

\FloatBarrier
\subsection{Design: capture law, models and prospective checks}
\label{sec:design-results}
\label{sec:law}
\begin{figure}[!ht]
\centering\includegraphics[width=\linewidth]{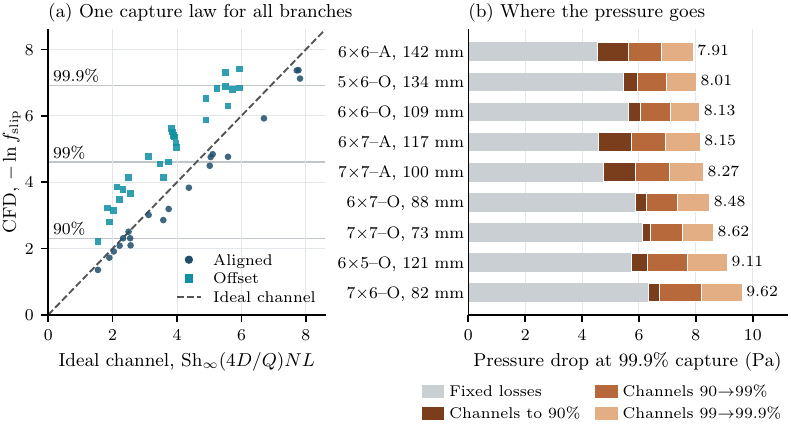}
\caption{The capture law and the pressure. (a) CFD slip against the slip of an ideal channel ($U=2$ m s$^{-1}$, design mesh). (b) Pressure drop at 99.9\% capture, split into the fixed losses and the channel pressure of each tenfold cut in slip.}
\label{fig:nines}
\end{figure}

The capture law (Eq.~\eqref{eq:slip}) says that the slip falls exponentially with the transfer area of the channels, $-\ln\fslip=\Sh\,(4D/Q)\,N\,(L+\Leq)$. An ideal channel, fully developed with a perfect-sink wall, has the fully developed value $\Sh_\infty=2.977$ of a square duct \cite{ShahLondon1978,Tronconi1992} and $\Leq=0$, so its slip is $-\ln\fslip=\Sh_\infty(4D/Q)NL$. Figure~\ref{fig:nines}a plots the slip of every converged nominal observation against this ideal value.
The aligned branches lie on or below the ideal line, consistent with maldistribution among channels, whereas the offset branches lie above it: their segment junctions add the capture of 24--32 mm of extra channel ($\Leq$). Table~\ref{tab:law} gives the fitted constants by architecture.

\begin{table}[!ht]
\centering\small
\caption{Constants of the capture law and of the pressure model fitted to the 14 branches ($U=2$ m s$^{-1}$, design mesh), as ranges by architecture; the pressure is $\Delta P=\Delta P_0+kL$.}
\label{tab:law}
{\adjustbox{max width=\textwidth}{\begin{tabular}{lccccc}\toprule
Architecture & Branches & $\Sh$ & $\Leq$ (mm) & $\Delta P_0$ (Pa) & $k$ (Pa m$^{-1}$)\\\midrule
Aligned (A) & 6 & 2.2--2.8 & 1--12 & 3.4--4.7 & 15--36\\
Offset (O) & 8 & 2.8--3.1 & 24--32 & 4.8--6.3 & 11--40\\
\bottomrule\end{tabular}
}}
\end{table}

Because the slip decays exponentially with length, every tenfold cut in slip, from 90 to 99\% or from 99 to 99.9\% capture, needs the same extra channel length within a geometry, $\ln10/[\Sh(4D/Q)N]$ from Eq.~\eqref{eq:slip}, which is 32--131 mm over the 14 branches. Each such step costs $k$ times that length, only 1.0--1.9 Pa, whereas the fixed losses $\Delta P_0$, which do not grow with length (ports, plenums and the segment junctions of the offset designs), carry most of the pressure at any capture (Fig.~\ref{fig:nines}b). The offset designs reach 90\% with a quarter to a half of the channel pressure of a later step, because their junctions already capture part of the vapour. At a given capture the geometry therefore changes the length, that is the contactor volume, far more than the pressure.

The law is useful only if it can be calibrated cheaply. Two CFD runs of different length fix the straight line of $\ln\fslip$ against $L$: its slope gives $\Sh$ and its intercept $\Leq$. To test this, the line is drawn through the two shortest runs of each of the 13 branches simulated at three or more lengths and used to predict the 20 longer runs left out of the fit: it misses the length they need by a median of 2.8 mm (Table~\ref{tab:calib}). One run cannot separate $\Sh$ from $\Leq$; with $\Leq=0$ the median error rises to 13.4 mm, and to 21.7 mm for the offset branches, whose $\Leq$ is 24--32 mm.

\begin{table}[!ht]
\centering\small
\caption{Error in length of the calibrated law on the 20 longer lengths held out of the fit, by number of calibration runs per branch: median and largest absolute error (mm). With one run $\Leq=0$.}
\label{tab:calib}
\begin{tabular}{lrrrrrr}\toprule
 & \multicolumn{2}{c}{Aligned} & \multicolumn{2}{c}{Offset} & \multicolumn{2}{c}{All}\\
\cmidrule(lr){2-3}\cmidrule(lr){4-5}\cmidrule(lr){6-7}
Runs & Median & Largest & Median & Largest & Median & Largest\\\midrule
One & 5.9 & 13.4 & 21.7 & 37.7 & 13.4 & 37.7\\
Two & 3.6 & 10.3 & 1.5 & 7.1 & 2.8 & 10.3\\
\bottomrule\end{tabular}
\end{table}

\begin{figure}[!ht]
\centering\includegraphics[width=\linewidth]{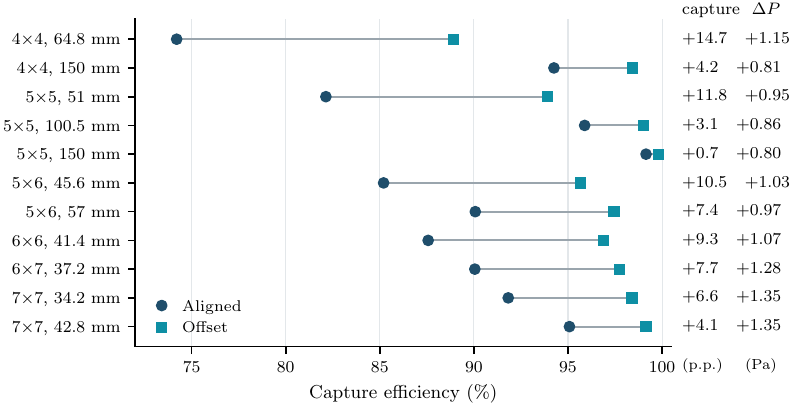}
\caption{The offset at equal length: capture of the aligned and offset designs at the 11 lengths simulated with both architectures ($U=2$ m s$^{-1}$, design mesh). At the right, the extra capture (percentage points) and extra pressure drop of the offset.}
\label{fig:offset-pairs}
\end{figure}

\label{sec:mechanism}
At equal length the offset captures more in all 11 pairs simulated with both architectures, for 0.8--1.4 Pa more (Fig.~\ref{fig:offset-pairs}). How much more depends on the vapour left: 10--15 percentage points where the aligned design captures 74--85\%, and less than one point at 99\%.

\begin{figure}[!ht]
\centering\includegraphics[width=\linewidth]{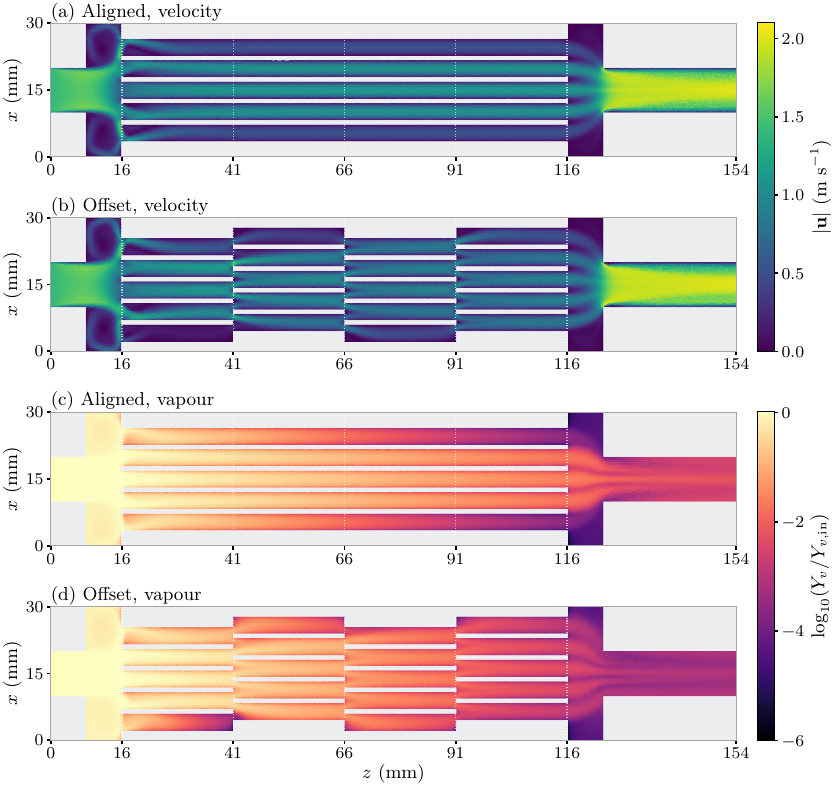}
\caption{CFD fields of the paired $5\times6$ designs ($L=100$ mm, 1.4 m s$^{-1}$, design mesh) on a longitudinal cut through one channel row ($y=12.6$ mm), flow from left to right; dotted lines mark the segment interfaces and grey regions lie outside the fluid. (a,b) Velocity magnitude; (c,d) vapour mass fraction relative to the inlet, on a logarithmic scale common to both architectures.}
\label{fig:cfd-longitudinal}
\end{figure}

\begin{figure}[!ht]
\centering\includegraphics[width=\linewidth]{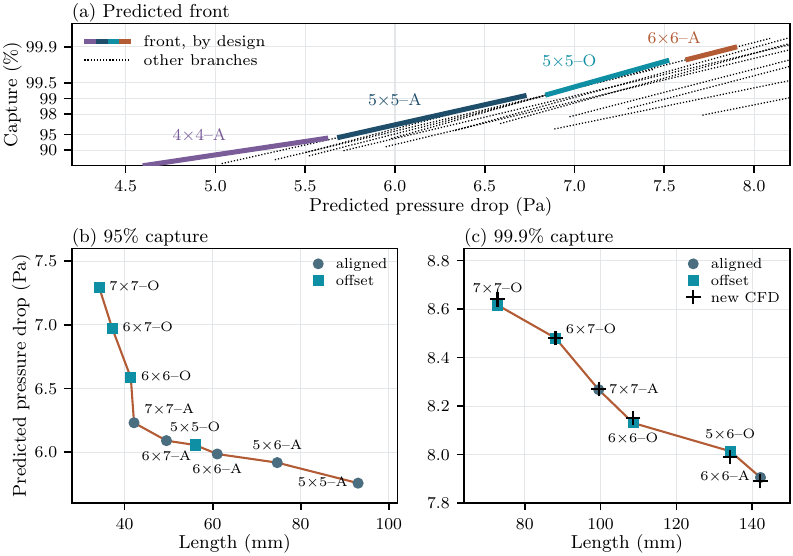}
\caption{Design screening with the calibrated laws ($U=2$ m s$^{-1}$, design mesh). (a) Lowest predicted pressure drop at each capture. (b, c) Nondominated designs at 95 and 99.9\% capture, predicted by the laws; crosses are new CFD at the predicted lengths.}
\label{fig:front}
\end{figure}

Near full capture the small gain still matters: at a fixed target it saves length, because each further point of capture costs more channel (Fig.~\ref{fig:front}). The fields of Fig.~\ref{fig:cfd-longitudinal} show where this extra capture comes from, in a paired comparison of $5\times6$--A and --O at 100 mm and 1.4 m s$^{-1}$. The inlet preferentially feeds the central channels, and the offset redirects the flow into new passages at each interface. The vapour profiles of the two architectures nearly coincide in the first segment and separate after the first interface, where the offset redistributes the flow among channels and shifts the concentration field by a fraction of a channel width, consistent with a concentration layer that restarts in each segment.

\label{sec:freedom}
Solving the $\varepsilon$-constraint problem of Eq.~\eqref{eq:design-screen} over the 14 fitted branches for a range of capture targets gives the predicted Pareto front (Fig.~\ref{fig:front}a). Figures~\ref{fig:front}b and c fix the target at 95 and 99.9\% and compare length and pressure: they show the designs that no other design beats on both.

Two conclusions follow from Fig.~\ref{fig:front}. First, the best geometry changes with the target. The front is formed by $4\times4$--A up to about 94\% capture, $5\times5$--A up to 99.1\%, $5\times5$--O up to 99.8\% and $6\times6$--A at 99.9\%. Coarse channels have the lowest friction and win while the capture is modest, and higher targets call for more transfer per unit length, so the channel count rises from 16 to 25 and then 36. The offset, which captures more at equal length (Fig.~\ref{fig:offset-pairs}), does not always pay for its extra pressure: aligned designs hold most of the front, and the offset gives the lowest pressure only from 99.1 to 99.8\%.

Second, the front is flat, so length, not pressure, decides. The fixed losses $\Delta P_0$ carry 57--73\% of the lowest pressure, and the nondominated designs span a factor of two to three in length for at most 27\% more pressure (Fig.~\ref{fig:front}b,c). The shortest designs are offset at both targets. At 95\% they already sit at the minimum length $12a$ and save at most 8 mm; at 99.9\% $7\times7$--O needs 73 mm, against 100 mm for $7\times7$--A and 142 mm for $6\times6$--A, for 0.4 and 0.7 Pa more. Smaller channels are limited in practice by clogging, coating thickness and manufacturing tolerance, which the gas-side model leaves to the designer.

Finally, new CFD tests the screen at 99.9\%. The six designs of Fig.~\ref{fig:front}c and $6\times7$--A were simulated at their predicted lengths, fixed before the runs: capture lies within 0.004 percentage points of the target and pressure within 0.3\% of the prediction (crosses in Fig.~\ref{fig:front}c). On the finer 0.4526 mm mesh the five refined designs need 1.5--2.7 mm more, at most 2\% of their length, and a margin of 4.8 mm on $6\times6$--A holds 99.90\% down to 0.36 mm. The calibrated law chooses geometry and length on the design mesh; the mesh adds a few millimetres.

\FloatBarrier
\subsection{Operation and control}
\label{sec:velocity}
\label{sec:humidity-control}
With a perfect-sink wall the vapour equation is linear, so capture efficiency does not depend on the inlet humidity and the captured rate is the efficiency times the vapour supplied. Three 3D CFD runs of the $6\times6$--A design at its final length of 147 mm and 2 m s$^{-1}$, at RH 20, 50 and 80\%, confirm it: the capture stays at 99.922--99.925\%, while the captured rate grows fourfold. The captured rate, not the efficiency, is therefore the variable to regulate, as in a field zeolite harvester whose daily yield followed absolute humidity \cite{Zeng2024}.

\begin{figure}[H]
\centering\includegraphics[width=\linewidth]{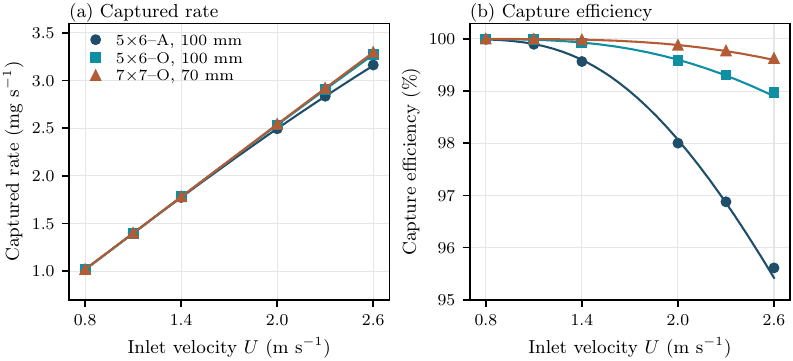}
\caption{Airflow as the handle in 3D: (a) captured rate and (b) capture efficiency of three designs against inlet velocity. Markers are CFD and lines a smooth fit through them.}
\label{fig:energy-map}
\end{figure}

The airflow, set by a fan or compressor at the inlet, is the variable that moves it. Over six inlet velocities from 0.8 to 2.6 m s$^{-1}$, a 3.25-fold increase in flow raises the captured rate 3.1--3.2-fold, because the efficiency falls only from above 99.99\% to 95.6--99.6\% (Fig.~\ref{fig:energy-map}a,b). The price is fan power: the hydraulic power grows 19--20-fold over the same range.

\begin{figure}[H]
\centering\includegraphics[width=\linewidth]{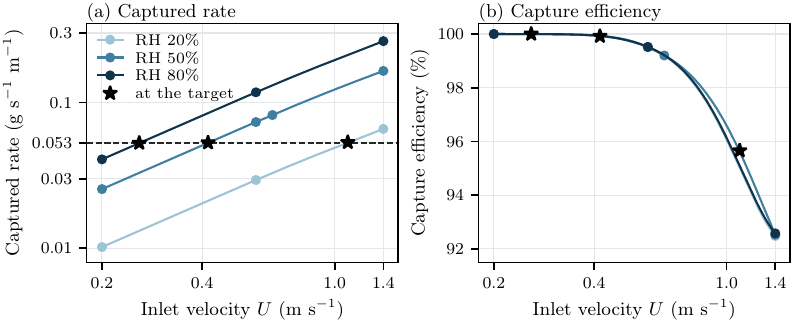}
\caption{Planar plant at RH 20, 50 and 80\%: (a) captured rate per unit depth and (b) capture efficiency. Circles are CFD joined by a shape-preserving interpolation; stars are steady runs at the velocities that hold the rate target (dashed).}
\label{fig:operability}
\end{figure}

\label{sec:2d-limits}
Testing the controller needs transient simulations, and a transient 3D trajectory costs roughly 80--145 times more than one on the planar plant of Section~\ref{sec:control-method}, a longitudinal section of $5\times6$--O. The planar plant responds to humidity as the 3D device does (Fig.~\ref{fig:operability}): from RH 20 to 80\% its efficiency curves coincide while its captured rate grows about fourfold. Holding the rate at its target of 0.053 g s$^{-1}$ m$^{-1}$ needs 1.10, 0.42 and 0.26 m s$^{-1}$ at RH 20, 50 and 80\%, confirmed by steady runs, and these points form the feedforward map. In dry air the higher velocity costs efficiency, 95.7\% at RH 20 against 99.9\% at RH 50, so a rate target and a minimum efficiency should be set together.

\begin{figure}[!ht]
\centering\includegraphics[width=\linewidth]{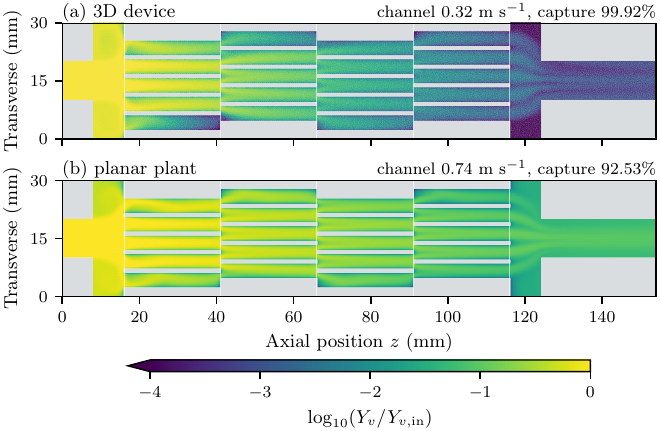}
\caption{Vapour mass fraction relative to the inlet for $5\times6$--O at 100 mm, 1.4 m s$^{-1}$ and RH 50\%: (a) 3D device on the cut at $y=12.6$ mm from which the planar section is taken, and (b) planar plant. Grey regions lie outside the fluid and white lines mark the segment interfaces; the 3D capture is integrated over the full outlet.}
\label{fig:dimensionality}
\end{figure}

The planar plant captures less than the 3D device. It omits the side walls and the three-dimensional redistribution of the flow, and at the same inlet velocity its channels run 2.3 times faster: at 1.4 m s$^{-1}$ it captures 92.5\% against 99.92\% in 3D (Fig.~\ref{fig:dimensionality}). It is therefore not used to predict 3D capture, but as a plant identified in its own right.

\begin{figure}[H]
\centering\includegraphics[width=.58\linewidth]{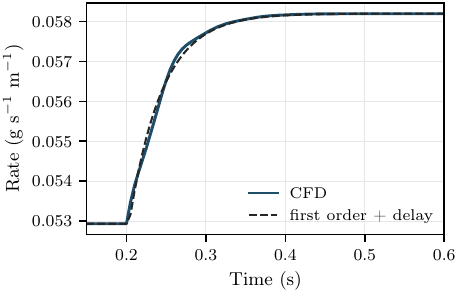}
\caption{Captured rate of the planar plant at RH 50\% after a +10\% velocity step at 0.2 s: CFD and first-order-plus-delay fit.}
\label{fig:tuning}
\end{figure}

The controller is tuned from one step test. A 10\% velocity step at RH 50\% gives a first-order response with a short delay (Fig.~\ref{fig:tuning}); halving the time step changes it by less than 6\% of its amplitude. The SIMC rules then give the PI controller of Table~\ref{tab:tuning}.

\begin{table}[H]\centering\small
\caption{Identified models (Eq.~\ref{eq:fopdt}) and controller parameters (Eqs.~\ref{eq:pi-tuning} and~\ref{eq:pi-ff}) at RH 50\% and $U_0=0.416$ m s$^{-1}$. Rates in g s$^{-1}$ m$^{-1}$, velocities in m s$^{-1}$.}
\label{tab:tuning}
\begin{adjustbox}{max width=\textwidth}\begin{tabular}{@{}lll@{}}\toprule
Element & Parameters & Source\\\midrule
Velocity path $G_p$ & $K_p=0.126$, $\tau_p=0.0415$ s, $\theta_p=0.0032$ s & 10\% velocity step\\
Humidity path $G_d$ & $\tau_d=0.058$ s, $\theta_d=0.042$ s & open-loop humidity step\\
PI controller & $K_c=7.35$, $T_i=0.0415$ s, $T_s=0.010$ s & SIMC, $\tau_c=\tau_p$\\
Feedforward $\mathcal F$ & delay 0.04 s, lead--lag $(0.0415s+1)/(0.058s+1)$ & $-G_d/G_p$\\
\bottomrule\end{tabular}\end{adjustbox}\end{table}

\begin{figure}[H]
\centering\includegraphics[width=\linewidth]{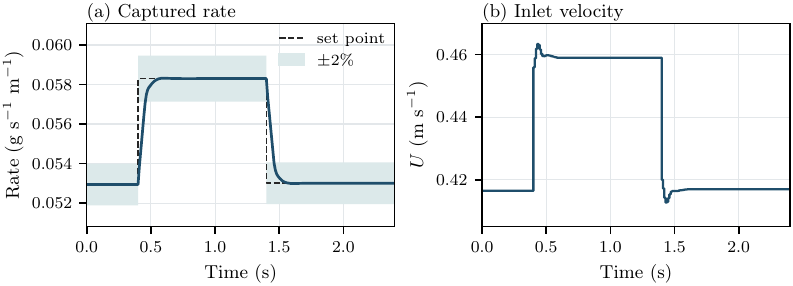}
\caption{Set-point change with the tuned PI on the planar plant at RH 50\%: (a) captured rate and (b) the inlet velocity it commands; shading is $\pm2\%$ of the set point.}
\label{fig:setpoint}
\end{figure}

A change of the rate set point tests the tuning: the rate enters the band of $\pm2\%$ of the set point, about a fifth of the 10\% step, within 0.06 s after the change and after its return, without saturating the velocity (Fig.~\ref{fig:setpoint}).

\begin{figure}[!ht]
\centering\includegraphics[width=\linewidth]{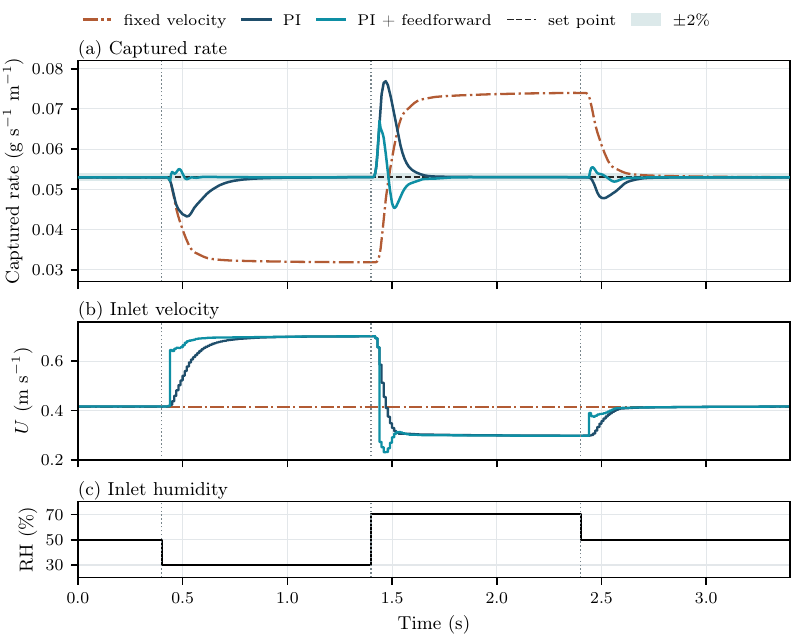}
\caption{Humidity steps on the planar plant with a fixed rate set point: (a) captured rate, (b) inlet velocity and (c) inlet humidity, at fixed velocity, with PI and with PI plus humidity feedforward (Eq.~\eqref{eq:pi-ff}), from the same initial state; shading is $\pm2\%$ of the set point.}
\label{fig:regulator}
\end{figure}

The humidity steps of Fig.~\ref{fig:regulator} then test disturbance rejection. At fixed velocity the rate drifts about 40\% from its set point and stays there. The PI raises the airflow as the humidity falls and lowers it as the humidity rises, and brings the rate back to the set point after every step; the integrated absolute error falls from 40.2 to 4.53 mg m$^{-1}$. The humidity feedforward changes the airflow so that its effect reaches the wall together with the humidity change, before the error builds up, and lowers the error to 1.37 mg m$^{-1}$, with a smaller peak after every step. On a second humidity sequence, with a ramp and a larger step and all parameters frozen, the feedforward again lowers the error relative to PI, by 72.4\%.

A unit of this geometry can thus hold its captured rate with the airflow alone over this humidity range. The humidity steps tested here are abrupt, whereas ambient humidity changes over minutes to hours; a controller that rejects these steps is therefore expected to follow the gradual changes of real operation as well.

\FloatBarrier
\section{Conclusion}
This work turns costly 3D CFD into a design, operation and control workflow for segmented monolith contactors that needs only a few simulations per geometry. A capture law fitted to a few CFD runs per geometry chooses geometry and length, new CFD at the chosen length confirms the choice, and a planar CFD section of a near-optimal geometry checks that it can be controlled. The wall is perfectly permeable to water vapour, so the results are the gas-side limit that any real sorbent approaches from below.

For design, the slip falls exponentially with length, so each tenfold cut in slip, from 99 to 99.9\% capture say, costs a fixed length for a given geometry, 32--131 mm over the 14 fitted branches. Two short runs per geometry calibrate this law to a median error of 2.8 mm in length. Seven designs predicted for 99.9\% capture and simulated with new CFD at lengths fixed in advance captured 99.896--99.902\%, and mesh refinement adds only a few millimetres to the length they need. Among the fitted geometries the best one changes with the target, from 16 channels at 90\% capture to 36 at 99.9\%.

Among these designs, length, not pressure, decides. In this laboratory-scale module the ports, plenums and offset junctions carry most of the pressure, so the nondominated designs span a factor of two to three in length for at most 27\% more pressure. The offset captures more than the aligned architecture at equal length, up to 15 percentage points where much vapour is left, for 0.8--1.4 Pa more; it seldom gives the lowest pressure, but it gives the shortest designs. At 99.9\% capture a $7\times7$ offset design of 73 mm, about half the length of the lowest-pressure design, captured 99.90\% in new CFD.

In operation, the captured rate, not the efficiency, follows humidity, and the airflow moves it: a 3.25-fold increase in flow raises the rate 3.1--3.2-fold while the efficiency falls only to 95.6--99.6\%. For control, on a planar CFD section of a near-optimal geometry, one step test tunes a PI controller of the captured rate that brings the rate within 2\% of a new set point in 0.06 s and rejects humidity steps that leave a sustained 40\% deviation at fixed velocity; a humidity feedforward lowers the integrated error by a further 70--72\%, also on a sequence not used to identify it. A unit of this geometry can thus be held at its rate target with the airflow alone.

For a designer, three rules follow:
\begin{itemize}\setlength{\itemsep}{1pt}\setlength{\topsep}{2pt}\setlength{\parsep}{0pt}
\item calibrate the capture law with two CFD runs (or measurements) per geometry and add a mesh margin of a few millimetres to the predicted length;
\item once the fixed losses dominate the pressure, choose among the designs by length, where the offset gives the shortest unit;
\item regulate the captured rate with the airflow, and set the rate target together with a minimum efficiency for dry air.
\end{itemize}

The same workflow carries over to the next steps: a finite-capacity, non-isothermal sorbent with regeneration and condensation, to predict breakthrough, cyclic water yield and energy demand; a transient 3D controllability check of the selected design; and the laboratory module, where measured responses can calibrate the capture law and tune the controller in the same way, with real sensors and actuators.

\section*{Data availability}
The data behind the figures and tables (steady 3D observations, fitted constants, mesh studies, prospective checks, and the steady maps and transient time series of the planar plant) are provided as CSV files with the supplementary material. Simulation case and data files, field states and analysis scripts are available from the corresponding author upon request.

\section*{Acknowledgements}
This work was financially supported by the Water4All -- Water Security for the Planet project. This research was also supported by Fundação para a Ciência e a Tecnologia, I.P./MECI through national funds: LSRE-LCM, UID/50020/2025 (\url{https://doi.org/10.54499/UID/50020/2025}); and ALiCE, LA/P/0045/2020 (\url{https://doi.org/10.54499/LA/P/0045/2020}). Margarida S.C.A. Brito acknowledges her contract with reference 2023.06416.CEECIND/CP2834/CT0008 (\url{https://doi.org/10.54499/2023.06416.CEECIND/CP2834/CT0008}) supported by the Individual Call to Scientific Employment Stimulus -- 6th Edition.

\section*{Declaration of generative AI and AI-assisted technologies in the manuscript preparation process}
During the preparation of this work, the authors used OpenAI Codex and Claude (Anthropic) to support writing, language editing and the organisation of the manuscript, with the aim of improving reproducibility and readability. After using these tools, the authors reviewed and edited the content as needed and take full responsibility for the content of the published article.

\bibliographystyle{elsarticle-num}
\bibliography{refs}
\end{document}